\documentstyle[11pt]{article}
 
\def\pa{\partial}

\def\g{\gamma} 
\def\a{\alpha}
\def\b{\beta}
\def\d{\delta} 
\def\e{\epsilon}

 \def\L{\Lambda}
\def\m{\mu}
\def\n{\nu}

\def\s{\sigma} 
\def\t{\tau}

\def\mn{{\mu\nu}}
\def\be{\begin{equation}}
\def\ee{\end{equation}}

\begin{document} \thispagestyle{empty} \begin{flushright}
\framebox{\small BRX-TH~532}\\
\end{flushright}

\vspace{.8cm} \setcounter{footnote}{0}

\begin{center} {\Large{\bf The Many Dimensions of Dimension}
    }\\[8mm]

S. Deser\\
Department of Physics, Brandeis University,\\ Waltham, MA 02454,
USA

{\small (\today)}\\[1cm]
\end{center}

\begin{abstract}
Some aspects of the history and impact of the dimension revolution
are briefly surveyed, starting with Nordstr\"{o}m's (1914) $D$=5
scalar gravity-electromagnetism unification.
\end{abstract}

\section{Introduction}

Calling into question the very 4-dimensionality of spacetime was
revolutionary and far-reaching:  one of the great innovations of
the 20${\rm ^{th}}$ Century.  That this was first done just nine
decades ago (albeit only after much hesitation over its first
fifty years) seems surprising, so quickly rooted has this concept
become. In this lecture I propose to survey -- necessarily very
briefly and unsystematically -- some of the stops on this evolving
journey, as one might explain them to the pioneers come back to
life. Consequently, I can hardly cover any of the more recent
flood of related ideas, let alone their mounting phenomenological
relevance.  That belongs to a different lecture altogether.  The
story here is primarily devoted to the period ending with the
early impact of strings and supergravity on the subject.  But
neither is this a historical lecture: priority and citation issues
can be found elsewhere than in the present impressionistic
excursion through our dimensional heritage.

Most conveniently, many of the older  original papers can be found
-- and translated into English -- in \cite{003}. The first book
covers Nordstr\"{o}m's and many later works as well; the second is
devoted to Oskar Klein's contributions. A very recent survey of
higher $D$, especially in the context of particle physics
phenomenology, is in \cite{005a}.  Finally, \cite{gibbons}
provides a current general analysis of consistency in dimensional
reduction methods, together with a history.

\section{History: Dimension from formal unification}

It is particularly appropriate to begin this brief voyage through
other dimensions with the pre-general relativity attempt, by
Nordstr\"{o}m \cite{001}, at formal unification of scalar gravity
and electromagnetism.  Although the gravity model was unrealistic,
its key insight -- dimension as unifier -- has become a standard
tool in all later attempts at finding ``theories of everything".

After 1905, it was clear that action-at-a-distance Newtonian
gravity was inconsistent with special relativity.  At the very
least it had to be covariantized, schematically:
\be
\nabla^2\phi = \rho \;\; \Rightarrow \;\; \Box \phi = T^\m_\m \; .
 \ee
 That the source of the wave equation is the trace of the matter
stress tensor is pretty well forced, since it must be a 4-scalar
that reduces to mass density in the nonrelativistic limit.  The
only other known force, electromagnetism, was represented by a
four-vector potential $A_\m$. Nordstr\"{o}m's insight was to
invent (1) and to unify\footnote{Of course, $A_\m$ was subject to
local gauge changes $\pa_\m \L (x)$, and $\phi$ only to constant
additions, but Nordstr\"{o}m emphasized the field strengths $(
F_\mn , \; F_{\m 5} = \pa_\m \phi )$.} the two forces into a
single 5-vector
\be
(A_\m , \phi ) \;\rightarrow \; B_M, \;\; M = (0,1,2,3,5)\; ,
 \ee
  in a
D=5 space.  As a result, the combined vector-scalar action of D=4
simply emerges from pure Maxwell in D=5, at least if one  omits
all $x^5$ dependence $(\pa/\pa x^5 = 0)$ {\it ab
initio}.\footnote{It was fortunately only discovered much later
(see \cite{gibbons}) that ``brutal" reductions to $D$=4 in which
unwanted fields and ``$x^5"$-dependence are simply dropped is not
always consistent!} That is,
\be
-\textstyle{\frac{1}{4}} \; F^2_{MN} \rightarrow -
\textstyle{\frac{1}{4}} \; F^2_\mn - \textstyle{\frac{1}{2}}
(\pa_\m \phi \pa^\m\phi ) g^{55}
 \ee
 Note the 55-signature here; its spacelike (+) sign is absolutely
 essential, but must be put in by hand,
 to keeping (scalar) gravity ghost-free\footnote{The long range common to
both forces was Nordstr\"{o}m's driving element, even though the
attraction-repulsion difference between them was not discussed.
{\it That} little problem goes back -- at least -- to Maxwell, but
the alternation of attraction/repulsion for even/odd spin
intermediary fields was not properly understood until QFT times!}
and -- consequently --
 attractive!
There is also an attendant (rather unnatural) source
``unification", with the 5-current $j^M$ defined as $(j^\m , \;
T^\a_\a)$.  Allowing some -- say periodic -- dependence on $x^5$
in this precursor model, would have similar effects to those
familiar from the Kaluza--Klein (KK) \cite{002} unifications to
which I now turn.

The Nordstr\"{o}m lesson was rediscovered (independently) by K and
K in the correct, Einstein gravity, context where the Maxwell
vector potential was leapfrogged by the metric tensor:
\be
(A_\m, \; g_{\a\b}) \rightarrow g_{MN} = (g_{\a\b}, g_{\m 5},
g_{55}) \; . \ee
 Here dominance by the highest spin now costs us an extra,
 scalar, field beyond the desired tensor-vector doublet.
 At linearized approximation $g_{MN} = \eta_{MN} + h_{MN}$, the
 action decomposes into the sum of spins 2, 1 and 0 parts:
 \be
 I_{KK} = \int d^5x \; \sqrt{-g} \; R(g_{MN}) \rightarrow
 I_{s=2}(h_{\a\b}) + I_{s=1}(h_{\a 5}) + I_{s=0}(h_{55}) \; .
 \ee
 All this is still in the $\pa/\pa x^5 = 0$ context
and there remains some unnaturalness about the source $T^{MN}$,
{\it e.g.}, how to identify $T^{55}$. Note that the correct choice
of 55 signature simultaneously ensures non-ghost behavior for both
$s$=1 and $s$=0.

Why did the pioneers stop at $D$=5?  The answer is Occam's razor:
there were only two known forces, so not only did D=5 suffice but
even it gave the (now superfluous) $g_{55}$ field, which however
prefigured later scalar-tensor gravity models. Indeed, as we shall
see, even now we are not sure we know what is the right $D$, if
that is even a correct question: the notion of dimension may
dissolve below Planck scales.

\section{Taking extra dimensions (really) seriously; $\mbox{\boldmath
$D$}$ from groups?}

As was especially emphasized by Klein's innovations, one should
not constrain higher $D$ models to be $x^5$-independent, or else
the extension -- however elegant -- is just a formal one. Indeed,
Klein was able to exploit $x^5$-dependence in quite ``modern"
ways, especially his derivation of electric charge quantization as
a result of assumed periodicity in $x^5$. Furthermore, he was
driven (albeit implicitly) to his now-famous ``almost"-derivation
(translated in \cite{003}) of Yang--Mills (YM) theory by extension
of these ideas. Ironically, it was only much later that a detailed
``KK" inclusion (in suitably higher $D$) of YM was formulated --
as a mere problem, in deWitt's 1963 Les Houches lectures
\cite{004} (also in unpublished work of Pauli). This notion has
been generalized to more abstract ``internal" groups that can
mimic real dimension through what can aptly be characterized as
``dimensional transmutation". Mixing of geometry and ``internal"
symmetries is clearly a fertile idea.

For us, dependence on ever higher dimensions, of whatever origins,
has become the norm -- we have much less respect for $D$=4 as the
``true" arena underlying dynamics, especially after supergravity
and superstrings (see below).

\section{Compactify: a lot/a little?}

We may not have much respect left for $D$=4, but we are still
constrained to live and observe there.  The direct presence of
extra dimensions must be sweepable under some (macroscopic) rug.
One way is making them very small and compactifying, as Klein
originally did for $x^5$.  [This is the usual ``garden hose"
picture of extra dimension in popular science expositions.]  More
recently (see \cite{005a}) there have been explorations of models
with relatively large -- almost macroscopic -- extra $D$, that
surprisingly cannot be excluded using observed gravitational
bounds. These remarks already point to the more general problem of
too much freedom: what is {\it the} vacuum state and how is it
fixed -- why is compactification at all (let alone narrowing down
to any specific one) the most attractive state?  Perhaps this is
another badly posed question -- are there many vacua, all of which
are of equal merit? In string theory, the plethora of Calabi-Yau
spaces allowed within a given $D$=10 is currently a similar
dilemma (see below).  Indeed, we don't even know in general why 4
big is ``better" than say 3 big, or all $D$ big! There is clearly
a lot of unknown territory even at the ``elementary" level.

\section{The quantum input: dimension as anomaly-killer}

Finding principles that actually fix the dimensionality of
space-time (and perhaps even its signature) is an old and as yet
unachieved goal.  However, one mechanism stands out in this quest,
because it is simultaneously (1) a clear application of quantum
field theory requirements, (2) leads to a unique dimensionality,
and (3) forms the basis of superstring theory.  The trick lies in
first transmuting $D$ into an index that counts the number of
excitations -- here scalar and spinor -- living in a lower -- here
$D$=2 -- spacetime.  As we all know, a (bosonic) string is a set
of one-dimensional objects evolving in time;  very roughly,
\be
I_s \sim \int \int d\s \, d\t \, \pa_\m \phi^a \, \pa_\n \phi^b \,
g_{ab} \, g^\mn \sqrt{g_2} \; , \;\;\; a,b = 1\ldots D \; .
 \ee
 The index $a$ represents the $\phi$ fields as the coordinates in a
``target space" with metric $g_{ab}$ and of dimension $D$, where
we live, while $\m$ ranges over the 2 dimensions $(\t ,\s )$ of
the world sheet on which each $\phi^a$ and the intrinsic 2-metric
depend. Now in QFT, classical symmetries, such as the conformal
invariance of a scalar field in two dimensions, best exhibited by
writing it in a gravitational background,
\be
I = -\textstyle{\frac{1}{2}}\int d^2x \; \sqrt{-g} \; g^\mn \:
\pa_\m \phi\pa_\n \phi\; , \;\;\; \d I=0, \;\;\; g_\mn \rightarrow
S (x) g_\mn \; , \ee can be, and usually are, broken (hence
``anomalies") by the necessity of introducing a regularization
scale or cutoff to make sense of the closed loops.  This
generation of anomalies is a very deep, nonperturbative,
phenomenon that has direct -- and observed -- physical
consequences.  A consistent quantum theory must be anomaly-free;
for the above bosonic string model, the coefficients of the
conformal anomaly add up to ($D$--26); bosonic strings have other
problems (such as tachyonic modes) however, that are absent in
their supersymmetric incarnations. These latter scalar $(\phi^a )$
plus spinor $(\psi^a )$, systems yield coefficients that add up to
($D$--10). Since we live in the target spaces, ours is a $D$=10
world.  Even if this number is slightly off our observed $D$=4, it
is one of superstring theory's impressive predictions: a unique
and not outlandish value of $D$, in addition to its other
wonderful (we hope!) attributes.\footnote{String theory is even
better served by anomaly-freedom than this: dangers from different
would-be anomalies reduce the number of possible strings to just
five, which in turn are linked by dualities into an essentially
unique model.}

\section{Embedding/Braneworlds}

Dealing with this enormous highly current topic can only be too
brief or too long, so I opt for the former (and refer to
\cite{005a} for more): once we are committed to a truly $D>4$
world, then our observed $D$=4 ``flatland" can have links to
itself through those other, $D$--4, dimensions: we live in a
submanifold rather than in a simple direct product, $4 \otimes
(D-4)$, space. Modern exploitations of this go under the rubric of
braneworlds, subspaces with considerably richer structure than
$D$=4 alone would allow.  For example, we would observe the usual
power laws $\sim 1/r$ for gravitational or electric potentials,
while significant deviations from them could be present ``nearby".
Normally, in $D$-spacetime, the Coulomb potential is $r^{-(D-3)}$,
yet there are various ways, besides a simple direct product, to
feel just $1/r$ in our particular corner.

\section{``Dynamical" Compactification}

The big questions we have already encountered in our survey is how
and why there are (at least at the present time) precisely $D$=4
macroscopic dimensions where our laws apply, and if one is greedy,
why our signature is $(-+++)$.  We have also noted that these
questions are still not only open, but possibly ill-posed! Here is
one possible way to distinguish $D$=4. Consider the popular value
of $D$, the context of supergravity (SUGRA), namely $D$=11, which
will be separately treated below. This value is the only
competitor to $D$=10; indeed one of the current active areas is
the relation of the string $D$=10 and the SUGRA $D$=11, evoking
the (mysterious) $M$-theory I will not mention further. Although
11 is not {\it a priori} more amenable than 10 to compactification
notions, there is one possible mechanism here \cite{005b}, that
illustrates possible attacks on the problem. We want to understand
why there is a 4+7 breakup, rather than say 7+4 (or any other)
with 4 big(ger) dimensions. Now one component of $D$=11 SUGRA is a
3-form potential $A_{MNP}$ and its attendant 4-form field strength
$B_{MNPQ} = \pa_{[Q}A_{MNP]}$. A caricature of the idea is that
some sort of symmetric vacuum breaking is caused by this totally
antisymmetric $B$-form, which thereby separates $D$=4 from the
rest, {\it i.e.}, $<B^{MNPQ}> \sim \e^{\mn\alpha\beta}$. That is,
the content of the system has built-in symmetry breaking potential
(just as Higgs actions do, in another story).    This suggestion
has the merit of showing how dynamics alone could dictate its own
``optimal" $D$; perhaps the breakup is even epoch-dependent and we
just happen to live in the 4+7 phase.

\section{SUGRA: Upper Bounds on $\mbox{\boldmath{$D$}}$}

Having gone from $D$=4 to ever higher dimensions, one begins to
wonder why there should be {\it any} upper bound to $D$ at all (we
will mention the limit $D\rightarrow\infty$ below), except in the
superstring case, where we encountered a precise value, $D$=10,
rather than merely a bound.  We now consider the very important
related area -- supersymmetry (SUSY) and SUGRA -- in which a whole
spectrum of dimensions is possible, but in which an upper bound is
in fact set by physical considerations of a down-to-earth kind. We
recall that SUSY is the study of Bose--Fermi matter systems,
including spins (0,$\frac{1}{2}$,1) while SUGRA deals with spin
3/2 and its spin 2 gravitational partner, possibly including lower
spins as well. In SUSY the symmetry is under constant (fermionic)
transformations, while SUGRA adds to general coordinate invariance
a sort of ``Dirac square-root" local fermionic parameter.  One of
the key requirements for a system to obey a supersymmetry algebra
is one-to-one matching of bosonic and fermionic degrees of freedom
(DoF).  This is what limits $D$, simply because the number of
tensors and of spinors grow very differently with dimension.
Roughly, the former grow as a power, $\sim D^2$, the latter
exponentially $\sim 2^D$.  Consider  first the original
\cite{006}, and simplest (``$N$=1") SUGRA, in $D$=4.  Here the
system is the sum of Einstein\footnote{There is also a version
incorporating a necessarily negative (Anti-deSitter) cosmological
term along with an (apparent) fermion mass term \cite{007}.}  and
massless spin 3/2 fields.  The graviton and fermion are separated
by a 1/2 unit of spin and rotate into each other; schematically
the transformation
$$
\d g_\mn = \a(x) (\g_\m\psi_\n + \g_\n \psi_\m ), \; \d\psi_\m =
D_\m \a (x)\;,
$$
where $\a (x)$ is a fermionic parameter, leaves the action
invariant.  In $D$=4 all massless particles have helicities $\pm
s$ only, {\it i.e.}, they all (except scalars) have 2 DoF, so the
numbers match. However, as $D$ grows, the number of Einstein
graviton DoF goes as $\frac{D(D-3)}{2}$, the number of
transverse-traceless components of the spatial ($D$--1) metric.
Instead, $\psi_\m$ is a (vector-)spinor and shares the $\sim 2^D$
(times the vector) growth of spinors. The only way to keep
bose/fermi DoF parity is to introduce more lower spin fields as
part of the supergravity multiplet, something that can also be
done at $D$=4, but as an option rather than necessity. One can
tabulate \cite{008} all fields' DoF at any $D$, so it is easy to
count the options in general. From this point of view, $D$=11 is
\cite{009} the highest dimension where the balance is kept without
making appeal either to spin $>2$ or to more than one graviton or
both. Now from the dynamical point of view, both of those latter
choices are forbidden: the coupling of gauge fields with $s>2$ to
gravity is inconsistent, and more than one gravity, {\it i.e.}, a
sum of two separate Einstein metric actions is -- as might be
expected -- not physical.  Hence it is simply ``ordinary"
dynamics, rather than the impossibility of a kinematical spin
matching in $D > 11$ that forbids SUGRA there, a very appealing
outcome.  Also, at $D$=11, there is but one way to balance the 44
gravitons in the bosonic sector to the 128 $\psi_\m$ components,
namely to add 84 3-form fields $A_{\mn\a}$. This uniqueness is one
the model's lasting attractions.\footnote{So unique is this
theory, in fact, that it is the one SUGRA, and indeed the one
known physical model, that does not even permit a cosmological
extension at all \cite{010}. It also forbids supermatter sources.}

\section{$\mbox{\boldmath$D \rightarrow \infty$}$?}

This is our shortest section.  One cannot avoid wondering what
sort of world would exist for $D\rightarrow\infty$.  There are
only two references \cite{011} to my knowledge, at least in the
context of Einstein gravity, with no very specific indication or
conclusion as yet; this may be a sign from heaven, to stick to
finite $D$, or perhaps we don't yet know how to extract the right
questions here either.

\section{$\mbox{\boldmath{$D$}}$ and Anthropic Principles}

The trend of our survey so far has been that there is no really
compelling theory of $D$ (except perhaps that from strings), nor
of the $D$=4+($D$--4) breakup for a given $D$, nor of the
signature of spacetime.  It is therefore worth mentioning a
currently fashionable idea, loosely styled anthropic: Let a
thousand vacua bloom; in the ensemble of all these possible
cosmologies, we're in the (only?) one that allows us.  In this
view, there is not point seeking some single string vacuum among
the immense multitude present in various Calabi--Yau spaces with
various, fluxes, brane choices, etc., provided that one of them is
ours! This statistical surrender to the counsel of despair may yet
be necessary, but it should perhaps not (yet) be welcomed. The
most recent, if hardly final, salvo in this battle is \cite{013}.

\section{$\mbox{\boldmath{$D<4$}}$: Laboratories of Real Physics}

We have, so far, explored $D>4$ worlds; what about dimensionally
challenged spacetimes with $D<4$?  Even $D$=1, while cramped, is
big enough for ordinary quantum mechanics:  $I = \frac{1}{2}\int
dt \: \dot{\mbox{\boldmath $r$}}^2(t)$ describes a $D$=1 base
manifold with a $D$=3 target space (or $D$=4 spacetime), while we
have seen that $D$=2 contains nothing less than string theory.
This leaves $D$=3, a dimension that contains real condensed matter
physics, as described by planar QED, enhanced by an extremely
relevant novel term, the Chern--Simons (CS)
invariant,\footnote{Historically, the first appearance of a CS
term was in the SUGRA context, via the 3-form of $D$=11.  There,
however it is trilinear, rather than quadratic, in the fields.}
\be
I_{CS} = \int d^3x \: \e^{\mn\a} A_\m \pa_\n A_\a = \int d^3x \:
A_\m \; ^*\!F^\m \ee
 where $^*\!F^\m$ is the (vector) dual of $F_{\n\a}$. Its
nonabelian counterpart is even more topologically endowed; they
are of particular interest also in the finite temperature context,
a rich and mathematically complex area.

While $D$=3 Einstein gravity (and SUGRA) are devoid of dynamics as
they stand, because Ricci and Riemann tensors are equivalent in
$D$=3,  yet here too (gravitational) CS terms exist and provide
true DoF when added to the Einstein action.  In both vector and
tensor settings cases, novel effects such as coexistence of
non-zero mass with gauge invariance, abound \cite{012}.  So $D$=3
can, in some ways, be regarded as the residue of a ``KK" $D$=4
manifold, with the CS terms descending from topological invariants
like $\int d^4x \: F_\mn \:^*\!F^\mn$, but that is a whole other
story.  One lesson for us, in any case, is the $D$=dependence of
interesting possible invariants: Not only do conventional actions
exist in various dimensions, but some do so in only one particular
$D$, others (like Gauss--Bonnet invariants) do so only above a
given $D$.

\newpage

 \noindent{\Large\bf Conclusion}
\vspace{.2in}

 In the preceding episodes, I have tried to sample,
in a very nonsystematic way, some of the legacy and promise of the
great concept that the $D$ of physical theories need not be the
same as that of the strictly $D$=4 world we seem to inhabit.  This
generalization fits the lesson of effective actions in physics:
the more encompassing a new level of physical law, the less its
concepts need resemble those of its limiting approximations.
Indeed, one could argue that the freeing of dimensionality may be
one of the most permanent new ``top-down" ideas created in the
past century: We are deeply indebted to its pioneers,
Nordstr\"{o}m, Kaluza and Klein.

This work was supported by the National Science Foundation under
grant PHY99-73935.


\begin{thebibliography}{999}
\bibitem{003}
{\it Modern Kaluza--Klein Theories}, T.\ Appelquist, A.\ Chodos
and P.G.O.\ Freund, eds. (Addison Wesley, 1987);{\it Oskar Klein
Memorial Lectures}, v1, G.\ Ekspong, ed. (World Scientific, 1991).
\bibitem{005a}
F.\ Feruglio ``Extra Dimensions in Particles Physics,"
hep-th/040133.
\bibitem{gibbons}
M. Cvetic, G.W. Gibbons, H. Lu and C.N.\ Pope, Class.\ Quant.\
Grav.\ {\bf 20} (2003) 5161-5194; G.W.\ Gibbons and C.N. Pope,
hep-th/0307052.
\bibitem{001}
G.\ Nordstr\"{o}m, Phys.\ Z.\ {\bf 15} (1914) 504.\bibitem{002}
Th.\ Kaluza, Verh.\ Preuss Ak.\ der Wiss (1921) 966; O.\ Klein,
Z.\ Phys.\ {\bf 37} (1926) 895; Nature {\bf 118} (1926) 516.
\bibitem{004}
B.S.\ deWitt in {\it Dynamical Theories of Groups and Fields}
(Gordon and Breach, NY (1965)) p.\ 139.
\bibitem{005b}
 P.G.O.\ Freund and M.A.\ Rubin, Phys.\ Lett.\ {\bf
97B} (1980) 233; F.\ Englert, {\it ibid}, {\bf 119B} (1982) 339.
\bibitem{006}
S.\ Deser and B.\ Zumino, Phys.\ Lett.\ {\bf 62B} (1976) 335; S.\
Ferrara, D.Z.\ Freedman and P.\ van Nieuwenhuizen, Phys.\ Rev.\
{\bf D13} (1976) 3214.
\bibitem{007}
P.K.\ Townsend, Phys.\ Rev.\ {\bf D15} (1977) 2802; S.\ Deser and
B.\ Zumino, Phys.\ Rev.\ Lett.\ {\bf 39} (1977) 1433.
\bibitem{008}
W.\ Nahm, Nucl.\ Phys.\ {\bf B135} (1978) 149.
\bibitem{009}
E.\ Cremmer, B.\ Julia and J.\ Scherk, Phys.\ Lett.\ {\bf 76B}
(1978) 409.
\bibitem{010}
K.\ Bautier, S.\ Deser, M.\ Henneaux and D.\ Seminara, Phys.\
Lett.\ {\bf 406B} (1997) 49;  S.\ Deser, L.\ Griguolo and D.\
Seminara, in preparation.
\bibitem{011}
A.\ Strominger, Phys.\ Rev.\ {\bf D24} (1981) 3082; N.E.J.\
Bjerrum--Bohr, hep-th/0305062.
\bibitem{013}
M.\ Dine ``Is There a String Theory Landscape: Some Cautionary
Notes," hep-th/0402101.
\bibitem{012}
S.\ Deser, R.\ Jackiw and S.\ Templeton, Ann.\ Phys.\ {\bf 140}
(1982) 372.
\end{thebibliography}
\end{document}